\documentclass[prb,aps,preprint,showpacs,floats]{revtex4}
\usepackage{mathrsfs}

\usepackage{amssymb}
\usepackage{graphicx}
\usepackage{graphics}

\begin{document}

\title{Melting of the vortex lattice in layered superconductors}
\author{Bo Feng, Zhigang Wu, and Dingping Li}
\thanks{corresponding author}
\email[E-mail address:]{lidp@mail.phy.pku.edu.cn}
\affiliation{Department of Physics, Peking University, Beijing
100871, China }

\begin{abstract}
The structure function of the vortex lattice of layered
superconductor is calculated to one-loop order. Based on a
phenomenological melting criterion concerning the Debye-Waller
factor, we calculate the melting line of the vortex lattice, and
compare our results to Monte Carlo simulation and experiment. We
find that our results are quantitatively in good agreement with
the Monte Carlo results. Moreover, our analytic calculation of the
melting line of BSCCO fits the experiment reasonably well in a
temperature  range not far from $T_{c}$.
\end{abstract}

 \maketitle
\section{INTRODUCTION}
Magnetic fields can penetrate into the layered high-$T_{c}$
superconductors(LHTS) and generate the well-known vortex
matter.\cite{Blatter} Thermal fluctuations and the effects of
disorder are able to drive the vortex matter to undergo very
complicated phase transitions among glass, solid and liquid phases.
\cite{Scheidl,Bhattacharya,Li} This led to a burst of investigations
both experimental and theoretical, to understand the physical
properties of such vortex matter.\cite{Bhattacharya} One interesting
aspect concerning vortex matter is to determine the phase transition
line of the melting between the vortex solid state and the vortex
liquid state.\cite{Nelson,Kierfeld} Several significant
experiments\cite{Zeldov,Liang,Willemin} have observed the phase
transition of the melting between the vortex solid state and the
vortex liquid state. In addition, magnetization jumps\cite{Zeldov,
Liang, Willemin, Nishizaki} and specific heat spikes\cite{Schilling,
Roulin, Bouquet} were observed, which indicate that the vortex
lattice melting is a first-order phase transition. An often used
theoretical description of the vortex lattice is the elastic
theory,\cite{Bhattacharya, Blatter, Brandt2,Houghton,Giamarchi}
which is based on the lowest energy excitations on a perfect vortex
lattice and can retain in fact most of the interesting physics. In
the framework of elastic theory, one uses a phenomenological
criterion, Lindemann criterion,\cite{Lindemann,Mikitik, Brandt1} to
calculate the phase transition line.

Near $T_{\rm c}$ vortices overlap and the elastic theory is
questionable. Furthermore for vortex liquid, the elastic theory is
not applicable. Another theoretical approach to study the phase
transition in vortex matter is studying the thermal fluctuations
of the more fundamental model, the Ginzburg-Landau(GL)
model.\cite{Tinkham} The model can describe very well both vortex
lattice and vortex liquid in the region near $T_{\rm c}$.

However, the full Ginzburg-Landau(GL) theory is too complicated, and
one needs to use some sort of approximations to advance the
theoretical investigation. Usually, the interested phase transition
is located not far away from $T_{\rm c}$. Near $T_{\rm c}$, it is
well known that one can use the lowest Landau level(LLL)
approximation.\cite{Abrikosov, Blatter} A number of researchers had
studied the vortex liquid phase.\cite{Ruggeri,Tesanovic1,Brezin1}
For solid phase, however, due to supersoft phonon modes, the
perturbation theory was questioned. The problem was resolved by
Rosenstein\cite{Rosenstein} and it was found that all infrared
divergencies are canceled. Non-perturbative Gaussian variational
calculation had been carried out.\cite{Li1} Spinodal line was
determined and recently was confirmed by experiments.\cite{Xiao,
Thakur} By comparing the  free energy of vortex liquid with vortex
solid, the melting line had been obtained.\cite{Li1} The result was
used to explain the melting transition of YBCO type (not too high
anisotropy).\cite{Willemin, Nishizaki} Recently it was also used to
explain the melting transition in low $T_{\rm c}$
materials.\cite{Kokubo}

While for highly anisotropic ones such as the BSCCO, the case is
far more complicated. The relevant model for highly anisotropic
superconductors is the Lawrence-Doniach-GL
model(LDGL).\cite{Lawrence}. The LDGL model has been proposed to
describe LHTS with weak Josephson interlayer
coupling.\cite{Blatter, Feinberg,Glazman,Zamora} However the full
studying of this model is not tractable.  We assume that we can
start from an effective LLL LDGL by integrating the higher Landau
level modes and focus our studies on this model. The phase diagram
of the LLL LDGL model had been investigated numerically in Ref.
\onlinecite{MacDonald}, and the melting transition line was
obtained. It is highly interesting if one can obtain the melting
transition of the LLL LDGL model analytically. However even for
the LLL LDGL, there is not a nonperturbative calculation of vortex
liquid energy, therefore we have not yet determined the melting
transition directly by comparing the liquid and solid free energy.

Therefore we come back to use more phenomenological criterions to
determine the melting transition. We might use the Lindemann
criterion to study the melting transition, however unfortunately
we do not know how to obtain the elastic modulii of the
Lawrence-Doniach model and used them to obtain the melting
transition.

In this paper, instead we will use another criterion to determine
the transition, the Debye-Waller factor criterion (DW
criterion).\cite{Ashcroft,Stevens} Based on this criterion, we
will study analytically the model and the result will be compared
\textbf{\emph{quantitatively }} with the Monte Carlo(MC) one in
Ref. \onlinecite{MacDonald}, and the model will also be used to
calculate the melting transition line in BSCCO.

The Debye-Waller factor is  the number of the original height of
the second Bragg peak (thermal fluctuation not considered) divided
by the height of the second Bragg peak with thermal fluctuation
considered.  Due to thermal fluctuations, the Debye-Waller factor
will be reduced. If the peak height is lowered to some extent, for
example, 60\%, the lattice will be melted. The applications of the
DW criterion in both Yukawa system\cite{Stevens} and three
dimensional (3D) anisotropic case of the high-$T_{c}$
superconductors\cite{Li1} indicate that the criterion is quite
accurate in determining the melting transition line. By using the
DW criterion, we find that our calculations fit the result very
well obtained by the MC simulation for studying the effective LLL
LD model. The analytic calculation of the melting transition line
is compared reasonable well with experiment.\cite{Beidenkopf} We
also stress that the calculations for the structure function and
the Debye-Waller factor are fairly simple and straightforward.

The paper is organized as follows. The model is described and a
perturbative mean-field solution is developed in section II. Then in
section III the structure function of the vortex lattice is
calculated to one-loop. A melting criterion is discussed in section
IV, comparison with MC simulations and experiment was also discussed
in this section.

\section{MODEL, MEAN FIELD SOLUTION, AND THE PERTURBATION THEORY}
\subsection{Model}
We start from the following Lawrence-Doniach free energy:
\begin{eqnarray}
F_{\rm GL}&=&d_{0}\sum_{n}\int
d^{2}\vec{r}\Bigg[\frac{\hbar^{2}}{2m_{ab}}(|D\psi_{n}|^{2})
+\frac{\hbar^{2}}{2m_{c}d^{2}}\nonumber\\
&&\times\left|\psi_{n+1}
-\psi_{n}\right|^{2}\nonumber
+a(T)|\psi_{n}|^{2}+\frac{b'}{2}|\psi_{n}|^{4} \Bigg],
\end{eqnarray}
where $\psi_{n}$ is the order parameter defined on discretely
labelled continuum layers, $d_{0}$ is the layer thickness, $d$ is
the interlayer spacing, the covariant derivative is defined by
${\bf D}\equiv \nabla-i(2\pi/\Phi_{0})$, and $\Phi_{0}\equiv
(hc/2e)$. In the limit that $d_{0}=d$ and $d$ goes to 0, the LD
model reduces to the 3D anisotropic Ginzburg-Landau model. For
layered superconductors far from $H_{c1}$ (this is the range of
interest in this paper), the magnetic field is homogeneous due to
the overlap of the vortices. We choose the Landau gauge ${\bf
A}=(By,0,0)$, which describes a nonfluctuating constant magnetic
field directed perpendicular to the layers.  For simplicity, we
assume $a(T)= -\alpha(1-t)T_{c}, t\equiv T/T_{c}$, and other
parameters are
temperature independent.%

For convenience, within the LLL approximation, we use the
following units to rescale the model:\cite{Li2} the units of
length of the ``ab'' plane and the ``z'' direction are
$\xi_{ab}=\sqrt{\hbar^{2}/(2m_{ab}\alpha T_{c})}$ and
$\xi_{c}=\sqrt{\hbar^{2}/(2m_{c}\alpha T_{c})}$, respectively; the
unit of magnetic field is $H_{c2}$, and the order parameter field
is rescaled as $\psi_{n}^{2}\rightarrow(2\alpha
T_{c}/b')\psi_{n}^{2}$. The dimensionless free energy in these
units is
\begin{eqnarray}
\frac{F_{\rm GL}}{T}&=&\frac{d_{0}}{\omega}\sum_{n}\int d^{2}\vec{r}
[\frac{1}{2}|D\psi_{n}|^{2}+\frac{1}{2d^{2}}|\psi_{n+1}-\psi_{n}|^{2}\nonumber\\
&-&\frac{1-t}{2}|\psi_{n}|^{2}+\frac{1}{2}|\psi_{n}|^{4}],
\end{eqnarray}
The dimensionless coefficient is $\omega=\sqrt{2{\rm Gi}}\pi^2 t$,
where the Ginzburg number is defined by ${\rm Gi}\equiv
\frac{1}{2}(32\pi e^{2}\kappa^{2}\xi T_{c}\gamma/c^{2}h^{2})^{2}$
and $\gamma\equiv\sqrt{m_{c}/m_{ab}}$ is the anisotropy parameter.

\subsection{Mean field solution}
By minimizing $F_{\rm GL}$ with respect to $\psi_{n}$, this standard
variation problem leads to the well-known GL equation
\begin{eqnarray}
\mathcal{H}\psi_{n}&+&\frac{1}{2d^{2}}(2\psi_{n}-\psi_{n-1}-\psi_{n+1})\nonumber\\
&-&a_{h}\psi_{n}+|\psi_{n}|^{2}\psi_{n}=0,
\end{eqnarray}
where $a_{h}\equiv (1-t-b)/2$, $b\equiv B/H_{c2}$, and
$\mathcal{H}\equiv-(D^{2}+b)/2$. If $a_{h}$ is sufficiently small,
the ${\rm GL}$ equation can be solved perturbatively. Within the LLL
approximation, one gets the mean field solution of the ${\rm GL}$
equation
\begin{eqnarray}
\psi_{n}=\Phi=\sqrt{\frac{a_{h}}{\beta_{A}}}~\varphi({\bf x}).
\end{eqnarray}
Where $\varphi({\bf x})$ is the Abrikosov's lattice
solution,\cite{Abrikosov} its definition is
\begin{eqnarray}
\varphi({\bf
x})&=&\sqrt{\frac{2\pi}{\sqrt{\pi}a}}\sum^{\infty}_{l=-\infty}
\exp\Bigg\{i\left[\frac{\pi l(l-1)}{2}+\frac{2\pi\sqrt{b}}{a}l
x\right]\nonumber\\
&&-\frac{1}{2}\left(y\sqrt{b}-\frac{2\pi}{a}l\right)^{2}\Bigg\},
\end{eqnarray}
and where $a=\sqrt{4\pi/\sqrt{3}}$,
$\beta_{A}\equiv\langle|\varphi|^{4}\rangle=\int_{cell}d^{2}x
|\varphi|^{4}(b/2\pi)\approx1.1596$ is the Abrikosov's constant, the
``cell'' here is a primitive cell of the vortex lattice. Obviously,
the mean field solution is independent of the layer index n.

\subsection{Fluctuation spectrum}
In order to get the excitation spectrum one expands the free energy
functional around the mean field solution.  The fluctuating order
parameter $\psi_{n}$ can be written as the sum of the mean field
part and a small fluctuating part
\begin{eqnarray}
\psi_{n}(x)=\Phi(x)+\chi_{n}(x).
\end{eqnarray}
We emphasize here that the argument ``$x$'' in (5) stands for a 3D
vector (i.e., $x=({\bf x}, x_{3})$), and the bold-face font (e.g.,
${\bf x}$) stands for the 2D vector in the ``ab'' plane in this
paper. The field $\chi_{n}$ can be expanded in a basis of
quasimomentum eigenfunctions ${\varphi_{\bf k}}$(within the LLL
approximation):
\begin{eqnarray}
\varphi_{\bf
k}&=&\sqrt{\frac{2\pi}{\sqrt{\pi}a}}\sum^{\infty}_{l=-\infty}
\exp\Bigg\{i\Bigg[\frac{\pi
l(l-1)}{2}\nonumber\\
&&+\frac{2\pi\left(\sqrt{b}x-\frac{k_{y}}{\sqrt{b}}\right)}{a}l-x
k_{x}
\Bigg]\nonumber\\
&&-\frac{1}{2}\left(y\sqrt{b}+\frac{k_{x}}{\sqrt{b}}-\frac{2\pi}{a}l\right)^{2}\Bigg\}.
\end{eqnarray}
In order to do the perturbation calculation more conveniently
(this can be seen later), we define $\beta_{{\bf k}}$ and
$\gamma_{{\bf k}}$
\begin{eqnarray}
&&\beta_{{\bf k}}=\langle|\varphi|^{2}|\varphi_{\bf k}|^{2}\rangle,\nonumber\\
&&\gamma_{{\bf k}}=\langle(\varphi^{\ast})^{2}\varphi_{-\bf
k}\varphi_{\bf k}\rangle,
\end{eqnarray}
while ${\bf k}=0$, $\beta_{0}, \gamma_{0}$ are shorted as $\beta,
\gamma$, respectively. We get
\begin{eqnarray}
\chi_{n}(x)=\frac{1}{\sqrt{2}}\int
d^{3}k\frac{e^{-ik_{3}nd}}{\sqrt{2\pi}}\frac{d_{\bf k}\varphi_{\bf
k}({\bf x})}{(\sqrt{2\pi})^{2}}(O_{k}+iA_{k}),
\end{eqnarray}
where $k_{1}, k_{2}\in [-\infty, +\infty]$, $k_{3}\in [-\pi/d,
\pi/d]$, $d_{\bf k}=\exp{[-i\theta_{\bf k}/2]}$ and $\gamma_{\bf
k}=|\gamma_{\bf k}|\exp{[i\theta_{\bf k}]}$. For simplicity, we have
used in (8) the ``real'' field $O_{k}$ and $A_{k}$, which satisfy
the relations: $O^{\ast}_{k}=O_{-k}$, $A^{\ast}_{k}=A_{-k}$. Within
the LLL approximation, at order $a_{h}$, the eigenstates are
$O_{k}$, $A_{k}$, We find that it is convenient for us to get the
eigenvalues by using $d_{\bf k}$ in the expansion of $\chi_{n}(x)$.
The eigenvalues are
\parbox{8.0cm}
{\begin{eqnarray*}
\epsilon_{O}&=&\tilde{\epsilon}_{O}+\frac{1}{d^{2}}(1-\cos{k_{3}d})\\
&=&a_{h}\left(-1+\frac{2}{\beta}\beta_{\bf k}+\frac{1}{\beta}
|\gamma_{\bf k}|\right)+\frac{1}{d^{2}}(1-\cos{k_{3}d}),\\
\epsilon_{A}&=&\tilde{\epsilon}_{A}+\frac{1}{d^{2}}(1-\cos{k_{3}d})\\
&=&a_{h}\left(-1+\frac{2}{\beta}\beta_{\bf k}-\frac{1}{\beta}
|\gamma_{\bf k}|\right)+\frac{1}{d^{2}}(1-\cos{k_{3}d}).
\end{eqnarray*}}\hfill\parbox{0.5cm}{\begin{eqnarray}\end{eqnarray}}
In particular, when $k\rightarrow 0$,
$\tilde{\epsilon}_{A}\approx0.1a_{h}|k|^{4}$, while
$\tilde{\epsilon}_{O}$ has a finite gap.

\section{STRUCTURE FUNCTION OF THE VORTEX LATTICE }
In this section we calculate the structure function to order
$\omega$ within the LLL approximation, i.e., neglecting higher
$a_{h}$ correlations. Firstly, we calculate the density-density
correlation function defined by
\begin{eqnarray}
\tilde{S}({\bf z},z_{3})=\langle\rho({\bf x},x_{3})\rho({\bf
x+z},x_{3}+z_{3})\rangle_{\bf x} ,
\end{eqnarray}
where $\rho(x)\equiv|\psi(x)|^{2}$, and the subscript ${\bf x}$ here
indicates the average over the unit cell. The correlation function
is calculated using the well-known Wick expansion:\cite{Zinn}
\begin{eqnarray}
\tilde{S}({\bf z},z_{3})=\tilde{S}_{\rm mf}+\omega \tilde{S}_{\rm
fluct},
\end{eqnarray}
where the first term is the mean field part, while the second term
is the correction due to thermal fluctuations.

\subsection{Mean field contribution} The mean field part is
\begin{eqnarray}
\tilde{S}_{\rm mf}=\langle|\Phi(x)|^{2}|\Phi(x+z)|^{2}\rangle_{\bf
x}.
\end{eqnarray}
The structure function is the fourier transform $S({\bf q},0)=\int
d{\bf z}\exp{[i{\bf q}\cdot{\bf z}]}\tilde{S}({\bf z},z_{3}=0)$,
hence, the mean field part of the structure function is
\begin{eqnarray} S_{\rm mf}({\bf q},0)&=&\int d{\bf z}\exp{[i{\bf
q}\cdot{\bf z}]}\langle|\Phi(x)|^{2}|\Phi(x+z)|^{2}\rangle_{\bf
x}\nonumber\\
&=&\left(\frac{a_{h}}{\beta_{A}}\right)^{2}\frac{b}{2\pi}\int d{\bf
y}e^{i{\bf q}\cdot{\bf y}}|\varphi({\bf y})|^{2}\nonumber\\
&&\times\int_{cell}d{\bf
x}e^{-i{\bf q}\cdot{\bf x}}|\varphi({\bf x})|^{2}\nonumber\\
&=&\left(\frac{a_{h}}{\beta_{A}}\right)^{2}4\pi^{2}\delta_{n}({\bf
q})\exp{\left[-\frac{q^{2}}{2b}\right]}.
\end{eqnarray}
In order to derive Eq.(13), we have used the following relation:
\begin{eqnarray}
&&\int_{A}d{\bf x} \varphi({\bf x})\varphi_{\bf k}^{\ast}({\bf
x})\exp[-i{\bf x}\cdot{\bf q}]\nonumber\\
&&=4\pi^2\delta_{n}({\bf q}-{\bf k})\exp\left[\frac{\pi
i}{2}(n_{1}^{2}-n_{1})\right]\nonumber\\
&&\times\exp\left[-\frac{\bf
q^{2}}{4b}-\frac{iq_{x}q_{y}}{2b}+\frac{ik_{x}q_{y}}{b}\right],
\end{eqnarray}
where $A$ is the sample area, and where we have used the notation:
$\delta_{n}({\bf q})\equiv\sum_{n_{1},n_{2}}\delta({\bf
q}-n_{1}\tilde{{\bf d}}_{1}-n_{2}\tilde{{\bf d}}_{2})$,
$n_{1}=(1/2\pi){\bf d}_{1}\cdot {\bf q}$, $n_{2}=(1/2\pi){\bf
d}_{2}\cdot {\bf q}$, $\tilde{{\bf d}}_{1}$, $\tilde{{\bf d}}_{2}$
are the reciprocal lattice basis vectors
\begin{eqnarray}
\tilde{{\bf d}}_{1}=\frac{2\pi\sqrt{b}}{a}\left(1,
-\frac{1}{\sqrt{3}}\right);~~~ \tilde{{\bf d}}_{2}=\left(0,
\frac{4\pi\sqrt{b}}{a\sqrt{3}}\right),
\end{eqnarray}
which are dual to the lattice basis vectors
\begin{eqnarray}
{\bf d}_{1}=\left(a/\sqrt{b},0\right);~~~ {\bf
d}_{2}=\left(a/2\sqrt{b},a\sqrt{3}/2\sqrt{b}\right).
\end{eqnarray}

\subsection{Fluctuation contribution}
We calculate the fluctuation contribution of the structure function
($S_{{\rm corr.}}$) to one loop. For convenience, the results are
divided into four parts:
\begin{eqnarray}
S_{{\rm corr.}}({\bf q},0)&=&S_{1}({\bf q},0)+S_{2}({\bf
q},0)\nonumber\\
&&+S_{3}({\bf q},0)+S_{4}({\bf q},0),
\end{eqnarray}
where $S_{1}({\bf q},0)$ is the fourier transform of
$\langle\Phi(x)\Phi(x+z)\chi^{\ast}_{n}(x)\chi^{\ast}_{n}(x+z)+
c.c.\rangle_{\bf x}$, $S_{2}({\bf q},0)$ is the fourier transform of
$\langle\Phi(x)\Phi^{\ast}(x+z)\chi^{\ast}_{n}(x)\chi_{n}(x+z)+
c.c.\rangle_{\bf x}$, $S_{3}({\bf q},0)$ is the fourier transform of
$\langle|\Phi(x)|^{2}|\chi_{n}(x+z)|^{2}+|\Phi(x+z)|^{2}|\chi_{n}(x)|^{2}\rangle_{\bf
x}$, and $S_{4}({\bf q},0)$ is the fourier transform of
$\frac{2a_{h}}{\beta_{A}}\langle|\varphi(x)|^{2}|\varphi(y)|^{2}\rangle_{\bf
x}(\nu_{1}^{2})$. We emphasize that the final term is due to the
vacuum renormalization, which cause the shift $\nu$ in
$\psi_{n}(x)=\nu\varphi_{n}(x)+\chi_{n}(x)$ be renormalized. To
one-loop order, let $\nu^{2}=\nu_{0}^{2}+\omega\nu_{1}^{2}$, here
$\nu_{0}^{2}=a_{h}/\beta_{A}$, and $\nu_{1}^{2}$ is given by
minimizing the effective one-loop free energy
\begin{widetext}
\begin{eqnarray}
-{\rm ln}Z=
\frac{L_{x}L_{y}L_{z}}{\omega}\left(-a_{h}\nu^{2}+\frac{1}{2}\nu^{4}\beta_{A}\right)
+\frac{1}{2}{\rm Tr\ln}\left[\tilde{\epsilon}_{O}({\bf
k},\nu)+\frac{1-\cos k_{3}d}{d^{2}}\right]+\frac{1}{2}{\rm
Tr\ln}\left[\tilde{\epsilon}_{A}({\bf k},\nu)+\frac{1-\cos
k_{3}d}{d^{2}}\right].
\end{eqnarray}
where $L_{x}, L_{y}, L_{z}$ are the scales of the sample. From
Eq.(18), we get
\begin{eqnarray}
\nu_{1}^{2}= \frac{-\sqrt{2}}{16\pi^{2}\beta_{A}}\int_{\bf k} \left[
\frac{2\beta_{\bf k}+|\gamma_{\bf k}|} {\sqrt{\tilde{\epsilon}_{O}
+\frac{d^{2}}{2}\tilde{\epsilon}_{O}^{2}}}+\frac{2\beta_{\bf k}
-|\gamma_{\bf k}|}{\sqrt{\tilde{\epsilon}_{A}
+\frac{d^{2}}{2}\tilde{\epsilon}_{A}^{2}}} \right]
\end{eqnarray}
Each term of the r.h.s of (17) is given as follows:
\begin{eqnarray}
&&S_{1}({\bf q},0)=\frac{\omega
a_{h}}{2\beta_{A}}\cos\left[\frac{k_{x}k_{y}+({\bf k}\times{\bf
Q})_{z}}{b}+\theta_{\bf k}\right]\exp\left[-\frac{q^{2}}{2b}\right]
\left[\sqrt{\frac{2}{\tilde{\epsilon}_{O}+\frac{d^{2}}{2}\tilde{\epsilon}_{O}^{2}}}
-\sqrt{\frac{2}{\tilde{\epsilon}_{A}+\frac{d^{2}}{2}\tilde{\epsilon}_{A}^{2}}}\right]\\
&&S_{2}({\bf q},0)=\frac{\omega
a_{h}}{2\beta_{A}}\exp\left[-\frac{q^{2}}{2b}\right]
\left[\sqrt{\frac{2}{\tilde{\epsilon}_{O}+\frac{d^{2}}{2}\tilde{\epsilon}_{O}^{2}}}
+\sqrt{\frac{2}{\tilde{\epsilon}_{A}+\frac{d^{2}}{2}\tilde{\epsilon}_{A}^{2}}}\right]\\
&&S_{3}({\bf q},0)=\frac{\omega a_{h}}{2\beta_{A}}\delta_{n}({\bf
q})\exp\left[-\frac{q^{2}}{2b}\right]\int_{\bf
k}\cos{\left[\frac{({\bf k}\times{\bf Q})_{z}}{b}\right]}
\times\left[\sqrt{\frac{2}{\tilde{\epsilon}_{O}+\frac{d^{2}}{2}\tilde{\epsilon}_{O}^{2}}}
+\sqrt{\frac{2}{\tilde{\epsilon}_{A}+\frac{d^{2}}{2}\tilde{\epsilon}_{A}^{2}}}\right]\\
&&S_{4}({\bf q},0)=-\frac{\omega a_{h}}{2\beta_{A}}\delta_{n}({\bf
q})\exp\left[-\frac{q^{2}}{2b}\right]\int_{\bf
k}\left[\frac{2\beta_{\bf k}+|\gamma_{\bf k}|}{\beta_{A}}
\sqrt{\frac{2}{\tilde{\epsilon}_{O}
+\frac{d^{2}}{2}\tilde{\epsilon}_{O}^{2}}} +\frac{2\beta_{\bf
k}-|\gamma_{\bf k}|}{\beta_{A}}
\sqrt{\frac{2}{\tilde{\epsilon}_{A}+\frac{d^{2}}{2}\tilde{\epsilon}_{A}^{2}}}\right],
\end{eqnarray}
where ${\bf q}={\bf k}+{\bf Q}$, ${\bf k}$ is the fractional part of
${\bf q}$, while ${\bf Q}$ is the integer part of ${\bf q}$. After
combining the mean field part and the one loop correction part of
the structure function, we get
\begin{eqnarray}
S({\bf q},0)=S_{{\rm mf}}+S_{{\rm corr.}}
=\left(\frac{a_{h}}{\beta_{A}}\right)^{2}4\pi^{2}\delta_{n}({\bf
q})\exp\left[-\frac{q^{2}}{2b}\right]+\frac{\omega
a_{h}}{2\beta_{A}}\exp\left[-\frac{q^{2}}{2b}\right][f_{1}({\bf
q})+\delta_{n}({\bf q})(f_{2}({\bf Q})+f_{3})],
\end{eqnarray}
where $f_{1}({\bf q}), f_{2}({\bf Q}), f_{3}$ are given as follows:
\begin{eqnarray} f_{1}({\bf
q})&=&\left[1+\cos\left(\frac{k_{x}k_{y}+({\bf k}\times{\bf
Q})_{z}}{b}+\theta_{\bf k}\right)\right]
\sqrt{\frac{2}{\tilde{\epsilon}_{O}({\bf k})
+\frac{d^{2}}{2}\tilde{\epsilon}_{O}({\bf k})^{2}}}\nonumber\\
&&+\left[1-\cos\left(\frac{k_{x}k_{y}+({\bf k}\times{\bf
Q})_{z}}{b}+\theta_{\bf k}\right)\right]
\sqrt{\frac{2}{\tilde{\epsilon}_{A}({\bf k})+\frac{d^{2}}{2}\tilde{\epsilon}_{A}({\bf k})^{2}}},\\
f_{2}({\bf Q})&=&\int_{\bf k}\left[-1+\cos\left(\frac{({\bf
k}\times{\bf Q})_{\rm z}}{b}\right)\right]
\left[\sqrt{\frac{2}{\tilde{\epsilon}_{O}({\bf
k})+\frac{d^{2}}{2}\tilde{\epsilon}_{O}({\bf k})^{2}}}
+\sqrt{\frac{2}{\tilde{\epsilon}_{A}({\bf k})+\frac{d^{2}}{2}\tilde{\epsilon}_{A}({\bf k})^{2}}}\right],\\
f_{3}&=&\frac{-1}{a_{h}}\int_{\bf k}
\left[\sqrt{\frac{2\tilde{\epsilon}_{O}({\bf
k})}{1+\frac{d^{2}}{2}\tilde{\epsilon}_{O}({\bf k})}}
+\sqrt{\frac{2\tilde{\epsilon}_{A}({\bf
k})}{1+\frac{d^{2}}{2}\tilde{\epsilon}_{A}({\bf k})}}\right].
\end{eqnarray}
It is very interesting to notice that each of the four terms
$S_{i}{(i=1,\cdots,4)}$ is divergent, respectively, as ${\bf
k}\rightarrow0$, however, the sums $S_{1}, S_{2}$ and $S_{3},S_{4}$
are not. Here we just take the sum $S_{1}+S_{2}$ as an example:
\begin{eqnarray}
S_{1}({\bf q},0)+S_{2}({\bf q},0)=\frac{\omega
a_{h}}{2\beta_{A}}\exp\left[-\frac{q^{2}}{2b}\right]f_{1}({\bf q}).
\end{eqnarray}
In order to see it more clearly, we use $\sqrt{b}$ to rescale the
momentum. As it can be shown that $k_{x}k_{y}+\theta_{\bf
k}=O(k^{4})$ when $k\rightarrow0$, the function $(k_{x}k_{y}+({\bf
k}\times{\bf Q})_{\rm z}+\theta_{\bf k})\rightarrow({\bf
k}\times{\bf Q})_{\rm z}$, and $1-\cos(k_{x}k_{y}+({\bf k}\times{\bf
Q})_{\rm z}+\theta_{\bf k})\rightarrow({\bf k}\times{\bf Q})_{\rm
z}^{2}$, hence it will cancel the $1/k^{2}$ singularity of
$\sqrt{2/(\tilde{\epsilon}_{A}+d^{2}\tilde{\epsilon}_{A}^{2}/2)}$.
It is also easy to show that $S_{3}+S_{4}$ is  the case.
Consequently, we get a not divergent result. The fluctuation
correction of the structure function (for non-peak region) is shown
in Fig.1.
\end{widetext}

\section{MELTING OF THE VORTEX LATTICE}
\subsection{A melting criterion}

The above calculations also indicate that thermal fluctuations will
reduce the intensity of the Bragg peak of the structure function. In
fact, the Debye-Waller factor has been used to describe the melting
of the lattice system.\cite{Stevens} If the intensity of the peak is
lowered to some extent, for example, 60\%, the lattice will be
melted. In 3D case, we know the exact melting transition temperature
via a different method.\cite{Li1} With this temperature, we find
that the one loop calculation of the Debye-Waller factor is reduced
to 50\%. This does not mean that the criterion ``60\%'' is wrong as
the higher order correction to the one loop calculation usually will
increase this value to some number above 50\% (it is too complicated
to calculate the structure function to two loop and we will leave it
to future studies).  In this paper, we also calculate the structure
function to one loop order. Thus we will use the "one loop
criterion" that the Debye-Waller factor calculated  to one loop is
reduced to 50\% at the melting transition line.

According to the definition of the Debye-Waller factor, we denote
the ratio of the one-loop value to the mean field value of the
intensity of the second Bragg peak by $\rho$, we get
\begin{eqnarray}
\rho=\frac{[f_{2}({\bf
Q_{1}})+f_{3}]\omega/2+4\pi^{2}a_{h}/\beta_{A}}{4\pi^{2}a_{h}/\beta_{A}},
\end{eqnarray}
where ${\bf Q_{1}}$ denotes the shortest reciprocal vector of the
triangular vortex lattice. We define the critical value of $\rho$
corresponding to melting by $\rho_{c}$. According to the above
discussion, the one loop criterion corresponding to $\rho_{c}$ is
about 50\%.

\subsection{Comparison with MC simulations}
Now we compare our results with MC simulations of the LLL layered
system in Ref. \onlinecite{MacDonald}. In Ref.
\onlinecite{MacDonald}, the authors use two dimensionless parameters
$g$ and $\eta$ to describe the melting transition of the system. The
$g$ and $\eta$ measure the intralayer and interlayer coupling,
respectively, and they are equivalent to the $t$ and $b$ in this
paper (in fact, they are the functions of $t$ and $b$). In order to
carry out the comparison, first, we make our notations consistent
with the ones of Ref. \onlinecite{MacDonald}.
\begin{eqnarray}
&&\alpha_{B}\equiv a(T)+\frac{\hbar e B}{m_{ab}c}=-2\alpha T_{c}a_{h},\\
&&g\equiv \alpha_{B}\sqrt{\frac{\pi \hbar c d_{0}\xi_{c}}{2\beta
k_{B} T e
B}}=-a_{h}\sqrt{\frac{\pi d_{0}}{b \omega}},\\
&&\eta\equiv\frac{\hbar^{2}}{2m_{c}(d\xi_{c})^{2}|\alpha_{B}|}=\frac{1}{2d^{2}a_{h}}.
\end{eqnarray}
Furthermore, from Eq.(26), (27), it is easy to see that $f_{2}({\bf
Q_{1}})+f_{3}$ is only dependent on $d^{2}a_{h}$, we define
$f(d^{2}a_{h})=f_{2}({\bf Q_{1}})+f_{3}$, then, we have
\begin{eqnarray}
f(d^{2}a_{h})=(\rho-1)\frac{8\pi^{2}a_{h}^{2}d_{0}}{\beta_{A} b
\omega d a_{h}^{1/2}}.
\end{eqnarray}
After combining Eq.(31),(32),(33) together, we get
\begin{eqnarray}
\frac{1}{\sqrt{2\eta}}f(\frac{1}{2\eta})=(\rho-1)\frac{8\pi
g^{2}}{\beta_{A}}.
\end{eqnarray}
According to the knowledge from 3D anisotropic model,\cite{Li1} we
choose $\rho_{c}=0.475$, hence, Eq.(34) gives out the relationship
between $g$ and $\eta$.

In Ref. \onlinecite{MacDonald}, the MC simulation was employed to
determine the melting transition. On the melting transition the
value of  $g$ is a function on $\eta$.  We denote  the function
$g$ in Ref. \onlinecite{MacDonald} by $g_{HM}$, and the $g$
determined by Eq.(34) is  denoted by $g$. The result of the
comparison is given in Table I. For typical LHTS such as BSCCO,
$\xi_{ab}$ is about 25{\AA}, $\gamma$ is about 200, $H_{c2}$ is
about 50T, $T_{c}$ is about 90K, and the interlayer spacing
$d_{0}$ is about 4{\AA}. For temperature and magnetic field  at
75K and 400G respectively, the  value of $\eta$ is about
$0.01$(actually $\eta$ on the all points on the theoretical  curve
on fig.2 is less than 0.1). when $\eta$ increases,  $|g|$ is
gradually larger than $|g_{HM}|$. As discussed in Ref.
\onlinecite{MacDonald} that the finite size effects of MC
simulation  become stronger as $\eta$ increased, and the finite
size effects lead the $|g_{HM}|$ to be less than its actual value.
In summary, we find the two results fit very well for not too big
$\eta$ (less than $0.1$) . This also demonstrates that the DW
criterion works well.
\begin{table}[]
  \centering
  \caption{Comparison of $g_{HM}$ with $g$}
  \vspace{.2cm}
\begin{tabular}{||c|c|c||}
\hline\hline
  $~~\eta~~$ & $~~g_{HM}~~$ & $~~g~~$ \\ \hline
  ~~0.005 ~~& ~~-5.2~~ & ~~-4.75~~ \\ \hline
  ~~0.01~~ & ~~-4.5~~ & ~~-4.36~~ \\ \hline
  ~~0.02~~ & ~~-3.9~~ & ~~-3.96~~ \\ \hline
  ~~0.06~~ & ~~-2.98~~ & ~~-3.31~~ \\ \hline
  ~~0.10~~ & ~~-2.74~~ & ~~-3.02~~ \\
 \hline\hline
\end{tabular}
\end{table}

\subsection{Comparison with experiment}
In Ref. \onlinecite{Beidenkopf}, the material parameters describe
BSCCO: $\kappa=100$, $\gamma=270$, $T_{c}=86$K, $d=15${\AA}. The
interlayer spacing $d_{0}$ and $H_{c2}$ have not given, we find
$d_{0}=4.1${\AA}, $H_{c2}=50$T give the best fit to the
experimental data, the results is shown in Fig.2.  The deviation
become large as $T$ reduces, this is expected as the effects of
disorder will be enhanced as temperature lowed. The effects of
disorder tend to lower the curve. However, the effects of thermal
fluctuations dominated in the region of our interest (near $T_{\rm
c}$). The comparison indicate that the effective LLL LD model is
quite good to describe the melting phase transition of the LHTS
near $T_{\rm c}$. In future work we will include the disorder
effect, and we expect that the result can be extended to the
region with lower temperature.

\section{CONCLUSIONS}
To conclude, we have calculated the structure function of layered
superconductors and the melting line can be obtained
quantitatively by "one-loop" DW criterion,i.e. the ratio of the
one-loop value of the intensity of the second Bragg peak of the
structure function to the mean-field value is about 50\%, the
solid melts. With this criterion, we calculate the melting line
and compare the results with existing MC results and experiments.
Our results fit the MC results very well. Moreover, our results
fit the experimental data reasonably well in the range not far
from  $T_{\rm c}$ (for BSCCO,  $T_{\rm c}=86$K, the range we
fitted is from $72$K to $86$K).

\begin{acknowledgements}
The authors acknowledge fruitful discussion with Prof. B.
Rosenstein and thank Prof. Zeldov for using his experimental data.
The work is supported by the National Natural Science Foundation
of China under Grant NO.90403002.
\end{acknowledgements}

\newpage

\textbf{Figure Captions}

\textbf{Figure 1}
Fluctuation correlation to the structure
function of the Abrikosov vortex lattice. The peaks at reciprocal
lattice points are removed, only the correction to the non-peak
region is plotted (i.e. only $f_{1}({\bf q})$ is plotted.

\textbf{Figure 2}

Comparison of the theoretical melting curve(line) of highly over
doped BSCCO with experimental results.
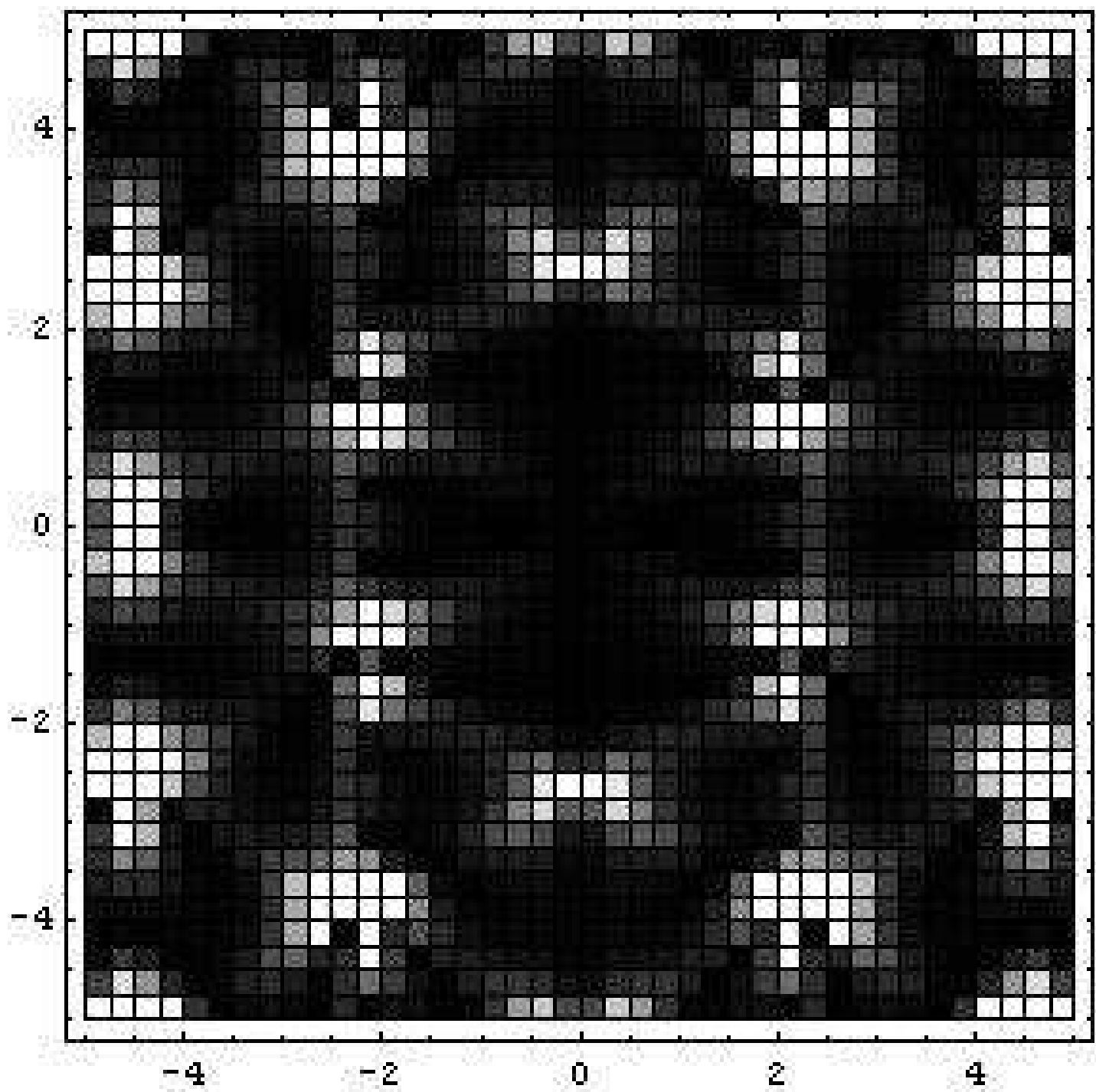
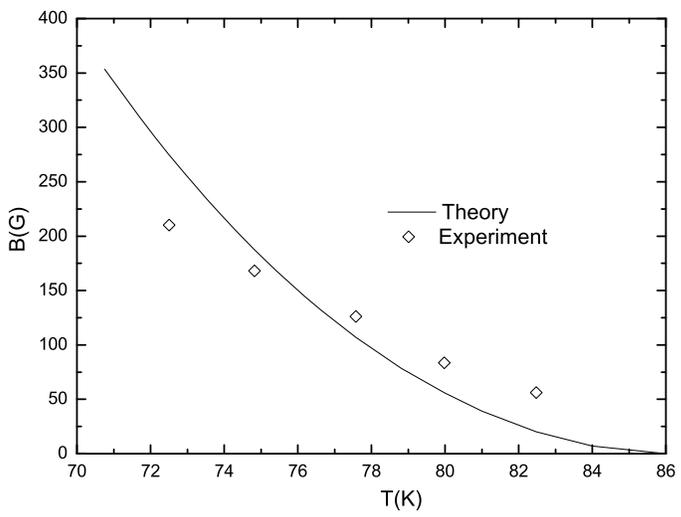
\end{document}